\documentstyle[12pt]{article}

\def\endpage{\vfill\eject}
\def\pbp{\langle\,\bar\psi\psi\,\rangle}

\newcommand{\AmS}{{\protect\the\textfont2
\renewcommand{\thesection}{\Roman{section}}
  A\kern-.1667em\lower.5ex\hbox{M}\kern-.125emS}}
 
\begin{document}
\rightline {\bf DFTUZ/97/13}
\vskip 2. truecm
\centerline{\bf Continuous Chiral Transition in Strongly Coupled}
\centerline{\bf Compact QED with the Standard Torus Topology}
\vskip 2 truecm
\centerline { V.~Azcoiti$^a$, G. Di Carlo$^b$, A. Galante$^{c,b}$, 
A.F. Grillo$^d$ and V. Laliena$^e$}
\vskip 1 truecm
\centerline {\it $^a$ Departamento de F\'\i sica Te\'orica, Facultad 
de Ciencias, Universidad de Zaragoza,}
\centerline {\it 50009 Zaragoza (Spain).}
\vskip 0.15 truecm
\centerline {\it $^b$ Istituto Nazionale di Fisica Nucleare, 
Laboratori Nazionali di Frascati,}
\centerline {\it P.O.B. 13 - Frascati 00044 (Italy). }
\vskip 0.15 truecm
\centerline {\it $^c$ Dipartimento di Fisica dell'Universit\'a 
dell'Aquila, L'Aquila 67100 (Italy)}
\vskip 0.15 truecm
\centerline {\it $^d$ Istituto Nazionale di Fisica Nucleare, 
Laboratori Nazionali del Gran Sasso,}
\centerline {\it Assergi (L'Aquila) 67010 (Italy). }
\vskip 0.15 truecm
\centerline {\it $^e$ Universit\"at Bern, Institute of Physics,}
\centerline {\it CH-3012, Bern (Switzerland).}
\vskip 3 truecm

\centerline {ABSTRACT}
\vskip 0.5truecm

\noindent
We analyze the phase diagram of compact QED on the torus 
with a chirally symmetric four fermion interaction.
Inside a mean field approximation for the four fermion term, a line
of first order phase transitions and another one 
of second order are found in the $(\,\beta, G\,)$
plane. Approaching the second order line 
a continuum limit can be defined. Critical exponents vary along 
this line in a similar way as in the non-compact model, suggesting that 
a non trivial interacting continuum theory can be constructed. 
  
\vfill\eject

\leftline{\bf 1. Introduction}
\vskip 0.5truecm

Non perturbative analysis of strongly coupled $QED$ started several years 
ago. The aim of these investigations was and is the search for non 
gaussian fixed points in the abelian gauge model; the main  
motivation for this kind of business being the relevance of 
non asymptotically free gauge theories in the context of composite models 
for the dynamical mass generation mechanism in the Standard Model.

First attempts to understand the nature of the continuum limit in full 
$QED$ with dynamical fermions \cite{PRIMERA} were done using the 
standard Wilson compact action for the pure gauge term. However there is 
no reason to prefer the compact formulation against the non compact one 
which, as well known, preserves also gauge invariance in the lattice 
formulation of the abelian model. Furthermore it was also argued that 
the non compact formulation, being free of monopoles as local minima 
of the action, would be nearest to the continuum formulation. This  
beside the early finding that the compact model suffers from a strong 
first order phase transition in the chiral limit \cite{KOGNOS} addressed 
the attention to the non compact formulation of the model 
where evidence for the existence of a non gaussian fixed point has been 
found \cite{VICENTE}.

There are also several more recent attempts to reexamine the old results 
on the first order character \cite{JERSAK,TRIO} of the phase transition 
in the pure gauge compact model. All these attempts are based in the 
suppression of monopoles well by modifying the standard Wilson action 
with the inclusion of a monopole term \cite{REBBI} or by changing the 
standard torus topology \cite{LANG,MARIANO}. In both approaches, which 
can be regarded as interpolations between the compact and non compact 
formulations but were the underlying pure gauge theory is not quadratic 
in the gauge field, the results suggest the existence of a 
non gaussian 
fixed point. Unfortunately no results with dynamical fermions are yet 
available here.

We want to show in this paper how the compact formulation of the abelian 
model with a chirally symmetric four fermion interaction 
has a line of continuous phase transition points, where a 
continuum limit can be defined. We will also show in the context 
of a mean field approximation for the fermion field that critical exponents 
take non mean field values, at least for critical points corresponding to 
not very small values of the bare gauge coupling.

The introduction of the four fermion term in the 
action is not an arbitrary attempt to modify the known results of the 
compact model. In fact there are strong arguments based on a non perturbative 
analysis of the Schwinger-Dyson equations in the continuum formulation
\cite{BARDEEN} suggesting that four fermion interactions become 
renormalizable outside perturbation theory. This is not surprising at all 
since naive dimensional arguments, which would exclude four fermion 
interactions as renormalizable, should not work near non gaussian fixed 
points where large anomalous dimensions can be generated.

The same analysis we are going to present here has also been done 
for the non compact formulation of the model \cite{GNJL}. As we will show, 
no qualitative differences in the phase diagram of both models appear 
in the weak gauge coupling regime, a region were we expect monopoles 
are dynamically suppressed. 

The gauged Nambu-Jona Lasinio $(GNJL)$ model regularized on a lattice 
and with compact gauge fields has the following action, 

$$ S = -\beta \sum_{n,\mu<\nu}
\cos\Theta_{\mu\nu}(n)\:+\:\bar\psi\Delta(\theta)\psi
\:+\:m\bar\psi\psi \:-\:
G\sum_{n,\mu}\bar\psi_n\psi_n\bar\psi_{n+\mu}\psi_{n+\mu} \eqno(1)$$

\noindent 
where $\beta$ is the inverse square gauge coupling, $\Delta(\theta)$ the
massless Dirac operator for Kogut--Susskind fermions, $G$ the four fermion
coupling and 

$$ \Theta_{\mu\nu}(n)\: = \: \theta_\mu(n) + \theta_\nu(n+\hat\mu) -
\theta_\mu(n+\hat\nu) - \theta_\nu(n) \eqno(2)$$

\noindent In the chiral limit ($m=0$) the action (1) is invariant under the
continuous transformations

$$\psi_n\;\;\rightarrow\;\;e^{i\alpha (-1)^{n_1+n_2+n_3+n_4}}\:\psi_n
\;\;\;\;\;\;\;
\;\;\bar\psi_n\;\;\rightarrow\;\;\bar\psi_n\: 
e^{i\alpha (-1)^{n_1+n_2+n_3+n_4}}
\eqno(3)$$

\noindent which define a $U(1)$ symmetry group. 

Since the four fermion interaction is not bilinear in the fermion field, 
it is not suitable for numerical simulations. To overcome this problem, 
we need to introduce an 
auxiliary vector field interacting with fermions, the path integral of 
which reproduces the four fermion coupling. To perform
numerical simulations of such a system is highly complex since it 
contains three parameters 
($\beta$,$m$ and $G$) and two vector fields \cite{EDIN}. Alternatively, 
we can use a mean field 
approximation to bilinearize the fermion action. Although quantitative
results will be affected by this approximation, we expect 
qualitative conclusions to be realistic \cite{GNJL}. 

As usual in the standard mean field technique, we make in (1) the following
substitution:

$$G\sum_{n,\mu}\bar\psi_n\psi_n\bar\psi_{n+\mu}\psi_{n+\mu}\;\;\rightarrow
\;\;2dG\langle\,\bar\psi\psi\,\rangle
\sum_n\bar\psi_n\psi_n
\eqno(4)$$

\noindent where $d$ is the space--time dimension. The action (1) becomes in this
way a bilinear in the fermion field and the path integral over the fermion 
degrees of freedom 
can be performed by means of the Matthews--Salam formula.
 In the following sections we shall analyze the phase structure and 
critical behavior of the model within this mean field approximation.

\vskip 1truecm 
\leftline{\bf 2. The Phase diagram}
\vskip 0.5cm

\par
The analysis of the phase diagram is similar to that of ref. \cite{GNJL},
and details can be found there. We shall study the chiral condensate
and the plaquette energy. 
The chiral condensate is an order parameter for chiral symmetry breaking. 
Within the mean field approximation (4) 
it is given by the selfconsistent equation: 

$$\langle\,\bar\psi\psi\,\rangle
=-2(m-8G\langle\,\bar\psi\psi\,\rangle)\,
\left\langle\,\frac{1}{V}\sum_{j=1}^{V/2}\frac{1}
{\lambda_j^2+(m-8G\langle\,\bar\psi\psi\,\rangle)^2}\,\right\rangle
\eqno(5)$$

\noindent where $V$ is the lattice volume and $\pm i\lambda_j$ 
$(j=1,\ldots,V/2)$
are the eigenvalues of $\Delta$, which depend on the compact 
gauge field $\theta$. The expectation value of the r.h.s. of (5) is taken over
the compact gauge field configurations and includes in the probability
measure, besides the exponential of the pure gauge
action, the fermionic determinant of standard QED evaluated at the 
effective mass $\bar{m} = m - 8 G \pbp$. 

In the chiral limit, eq. (5) becomes

$$\langle\,\bar\psi\psi\,\rangle=
16G\langle\,\bar\psi\psi\,\rangle\,
\left\langle\,\frac{1}{V}\sum_{j=1}^{V/2}\frac{1}
{\lambda_j^2+64G^2{\langle\,\bar\psi\psi\,\rangle}^2}\,\right\rangle
\eqno(6)$$
 
\noindent 
This equation is always verified if $\pbp=0$, and this is the only
solution in the symmetric phase. In the broken phase, where $\pbp\neq 0$,
the chiral condensate is given by the equation: 

$$1=16G\,\left\langle\,\frac{1}{V}\sum_{j=1}^{V/2}\frac{1}
{\lambda_j^2+64G^2{\langle\,\bar\psi\psi\,\rangle}^2}\,
\right\rangle
\eqno(7)$$

\noindent 
the solution of which minimizes
the free energy. It is therefore the physical (stable) solution.  
In the $G\rightarrow\infty$ limit the solution of eq. (7) at 
leading order is

$$1=16G\frac{1}{64G^2{\langle\,\bar\psi\psi\,\rangle}^2}\;
+\;O(\frac{1}{G^2})
\eqno(8)$$

\noindent This shows that at large $G$ chiral symmetry is spontaneously
broken for all values of $\beta$, the order parameter being   

$$ \langle\,\bar\psi\psi\,\rangle\sim\frac{1}{2\sqrt{G}}
\eqno(9)$$

The existence of a second order phase transition line separating 
the broken and symmetric phases follows
from eq. (7) \cite{GNJL}. 
It gives the following expression for the critical line:

$$G_c(\beta)=\frac{1}{\frac{16}{V}\,\langle\,\sum_{j=1}^{V/2}\frac{1}
{\lambda_j^2}\,\rangle}
\eqno(10)$$

\noindent 
where the probability distribution function in the previous mean value 
is that of massless compact $QED$.
The r.h.s. of eq. (10) is proportional
to the inverse transverse chiral susceptibility of compact QED 
in the chiral limit, 
which can be obtained from the identity

$$ \chi_T\:=\: \lim_{m\rightarrow 0}\,\frac{\pbp}{m} \eqno(11) $$

\noindent This allows us to discuss qualitatively the phase diagram of
the compact $GNJL$ model from the known properties of compact QED.

First, let us remember the main features of compact QED. 
In the chiral limit a strong first order phase transition
takes place at a value of the coupling $\beta_c\sim 0.9$ \cite{MFA}. 
It separates 
a strongly coupled phase, $\beta<\beta_c$, where
chiral symmetry is spontaneously broken, from a Coulomb phase,
$\beta>\beta_c$, characterized by a symmetric vacuum. 
At $\beta=\beta_c$ the chiral 
condensate shows a jump from zero to some non--zero value, and 
also the plaquette energy jumps. 
This first order phase transition
continues at non--zero bare fermion mass \cite{KOGNOS,COMQED}
along a line in the $(\beta, m)$
plane which connects the $(\beta_c, 0)$ point in the chiral limit with
the $(\beta_q, \infty)$ transition point of the quenched theory, 
at $m=\infty$ ($\,\beta_q\sim 1\,$, \cite{TRIO}). 
Both the chiral condensate and plaquette energy are 
discontinuous along this line \cite{MFA}. We denote this line 
by $(\,\beta, m_c(\beta)\,)$. 

Now, let us switch on the four fermion interaction. For $\beta$ values 
larger than $\beta_c$ the transverse susceptibility of compact QED is finite
in the chiral limit, and we get a line of second order phase transitions 
eq. (10) which
separates a phase where chiral symmetry is spontaneously broken, when
$G > G_c(\beta)$, from a symmetric phase, when $G < G_c(\beta)$.
When $\beta$ approaches $\beta_c$ from the right hand side, the $QED$  
massless transverse
susceptibility remains finite, and thus $G_c(\beta_c) >0$. 
This is a significant difference respect to the non compact model where, 
due to the fact that the susceptibility diverges in the 
$\beta\rightarrow\beta_{c}^+$ limit, $G_c(\beta_c) =0$.

In the broken phase of compact QED the transverse 
susceptibility is divergent in the chiral limit 
due to the Goldstone theorem, eq. (11). Therefore chiral symmetry is 
spontaneously
broken for all values of $G$ if $ \beta <\beta_c$. On the vertical line
$\beta=\beta_c$, $0\leq G < G_c(\beta_c)$ the transition is first order 
since in this range of $G$ the chiral condensate jumps from a non 
vanishing value at $\beta<\beta_c$ to zero at 
$\beta > \beta_c$. The plaquette energy is also discontinuous along this 
line.

The first order line at $m\neq 0$ in compact QED results 
in the chiral limit of the 
compact $GNJL$ model in a first order phase transition line which prolongates 
the vertical line from the point $(\,\beta_c, G_c(\beta_c)\,)$, ending 
at $\beta_q$, 
$G=\infty$. These first order transitions are originated by the effective
fermion mass $\bar{m}$ generated when $\pbp\neq 0$. Since the plaquette
energy and the chiral condensate for the compact $GNJL$ model in this mean
field approximation are obtained from those of compact QED evaluated
at an effective fermion mass $\bar{m}$, it is clear that for all
$\beta_c < \beta < \beta_q$ there exists a $G_{fo} > G_c(\beta_c)$ such that 

$$ -\, G_{fo} \pbp\,(\beta, G_{fo}) = m_c(\beta) \eqno(12) $$

\noindent 
with $G_{fo}(\beta_q)=\infty$. On this line 
$(\,\beta, G_{fo}(\beta)\,)$
the plaquette energy and the chiral condensate show a jump, 
producing a first order phase transition. In Figure 1 we show a qualitative 
picture of the phase diagram for the compact $GNJL$ model.

\vskip 1truecm
\leftline{\bf 3. Critical exponents}
\vskip 0.5truecm

\par
Since second order phase transitions are usually related to divergent 
correlation lengths, we expect that a continuum limit can be 
defined along the critical line given by eq. (10). 
The nature of the renormalized theory will depend on the critical
exponents at each critical point. 
The calculation of the critical exponents in this compact model is 
analogous to that of the noncompact version \cite{GNJL}. Here we will 
briefly comment the way to calculate them and refer the reader to ref.
\cite{GNJL} for details.

We write the chiral condensate of compact QED (without the four fermion
coupling) as

$$\langle\,\bar\psi\psi\,\rangle\:=\: -\,2\,m\, F(\beta,m)
\eqno(13)$$

\noindent
where $F$ is a function of $\beta$ and $m$, whose behavior when
$m\sim 0$ is

$$F(\beta,m)\:=\: F(\beta,0) \:+\: B\,m^\omega\:+\:\ldots
\eqno(14)$$

Equation (13) evaluated at the effective fermion mass 
$\bar{m} = m - 8 G \pbp$ is the equation of state for the $GNJL$ model 
in our mean field approach. From (13) and (14) and after simple algebra 
\cite{GNJL} we get the following relations for the critical exponents: 

$$ \delta = \omega + 1 \;\;\;\;\;\;\;\;\;\;\;\;\;\;\;\;\;\;\:\: \beta_m = 
\frac{1}{\omega} \;\;\;\;\;\;\;\;\;\; \gamma = 1 \eqno(15)$$

The practical rule to measure the critical exponents in the $GNJL$ model 
at a given critical point $(\beta, G(\beta))$ is therefore to measure 
the $\omega$ exponent of the subleading contribution to the chiral 
condensate in the Coulomb phase of compact $QED$. This exponent 
can be related to the behavior at the origin of the spectral 
density of eigenvalues of the Dirac operator. In fact, writing

$$ \pbp = - \int_0^\infty \,d\lambda\,\rho(\lambda)
\frac{2 m}{\lambda^2 + m^2} \eqno(16) $$

\noindent
the following relation is derived:

$$\omega = p-1\;\;\;\;(\,1<p\leq 3\,)\;\;\;\;\;\;\;\;\;\;\;
\omega = 2 \;\;\;\; (\,p>3\,) \eqno(17)$$

\noindent
where $p$ in (17) is the exponent which controls the behaviour of 
$\rho(\lambda)$ near the origin, i.e., $\rho(\lambda)\sim\lambda^p$ 
when $\lambda\rightarrow 0$. Notice that $p$ is always larger than 1 in 
the Coulomb phase of $QED$ as follows from a finite massless 
susceptibility.

The value of $p$ can be obtained from the analysis of what we call 
generalized susceptibilities, defined by \cite{GNJL}

$$\chi_{q} \:=\: \frac{1}{V}\left\langle\,\sum_{j=1}^{V/2}\frac{1}
{\lambda_j^q}\,\right\rangle.
\eqno(18)$$

\noindent
They are finite if $q < p+1$ and divergent otherwise. On finite lattices
we expect 

$$\chi_{q} \:\sim\: A_q \:+\: B_q\, L^{\alpha(q-p-1)} \eqno(19)$$

\noindent
with a logarithmic divergency with the lattice size in the especial case 
$q=p+1$. $A_q$ and $B_q$ in (19) are constants independent 
of the lattice size and $\alpha$ is some positive number. 
These finite size scaling relations 
allow us to calculate $p$ and therefore the critical exponents. 

The procedure we use is as follows. 
From simulations of compact QED in the chiral limit using the MFA approach 
\cite{MFA}, we are able to compute the generalized susceptibilities for
several values of $q$ and lattice size $L$. Fitting them with 
the function 

$$\chi_{q}^{-1}(L) = a_q+b_q L^{-c} \eqno(20)$$

\noindent
it is possible to extrapolate the results to the thermodynamical limit.
We obtain in this way the critical value $q_c$ such that $a_q > 0$ for 
$q < q_c$ and $a_q = 0$ for $q > q_c$. Eq. (19) implies $p = q_c - 1$, and
from (17) we get 

$$ \delta = q_c - 1\;\;\;\;\;\;\;\;\;\;\;\;\;\;\;\; 
\beta_m = \frac{1}{q_c - 2} \eqno(21) $$
 
We have applied this procedure to two points on the critical line of 
Fig. 1, $\beta=1.020, G=0.078$ and $\beta=1.045, G=0.099$. 
The corresponding critical 
values of $q$ have been obtained from the finite size scaling analysis 
(20) on $4^4,6^4,8^4$ and $10^4$ lattices. The extracted $q_c$ 
values together the corresponding critical exponents are reported in 
Table I. As in the non compact $GNJL$ model \cite{GNJL}, critical indices 
far from their mean field values are obtained for $\beta$ values near 
the phase transition point of $QED$.

\begin{table}[h]
\centering
\begin{tabular}{|c|c|c|c|c|c|}
\hline
\ $\beta_c$ & $G_c$ & $q_c$ & $\beta_m$ & $\delta$ & $\gamma$  \\
\hline
\hline
\ 1.020&0.078(1)&1.5(2)&2.0(6)&1.5(2)& 1 \\ 
\hline
\ 1.045&0.099(3)&2.7(2)&0.59(7)&2.7(2)& 1 \\ 
\hline
\end{tabular}
\end{table}

\vskip 1truecm
\leftline{\bf 4. Discussion}
\vskip 0.5truecm

\par
We have derived in this paper the phase diagram of the compact $GNJL$ model. 
A first order line starting from the critical coupling of the $QED$ model 
appears for every value of the four fermion coupling, ending at the critical 
value of quenched $QED$ at $G=\infty$. What is more exciting however is 
the finding of a second order line in the compact formulation of the model. 
This continuous phase transition (discontinuous line in Fig. 1) prolongates 
to the weak gauge coupling region approaching the $\beta=\infty$ axis at 
some finite value of $G$, a region where certainly monopole condensation 
or monopole percolation does not hold. 

We have also shown how critical exponents, computed in the mean field 
approximation for the fermion field but taking into account fluctuations 
of the gauge field, take non mean field values and change along the 
critical line. This is very similar 
to what we found in the non compact version of the model \cite{GNJL}. 
Again in the compact formulation, fluctuations of the gauge field seem 
to play a fundamental role in driving critical exponents to non 
gaussian values. One is then tempted to speculate that both formulations 
would drive to the same renormalized continuum physics. 

The last but not less important feature of the compact formulation of the 
gauged Nambu-Jona Lasinio model is the relevance of the four-fermion 
coupling in the search for a continuum limit. In contrast to what happens 
in the non compact formulation the four-fermion coupling, that as stated 
before becomes renormalizable outside perturbation theory, is an essential 
ingredient to define a continuum limit for the compact model.

\endpage

\vskip 1 truecm
\leftline{\bf Figure captions}
\vskip 1 truecm
{\bf Figure 1.} Phase diagram of the compact GNJL model in the 
$(\,\beta, G\,)$ plane.

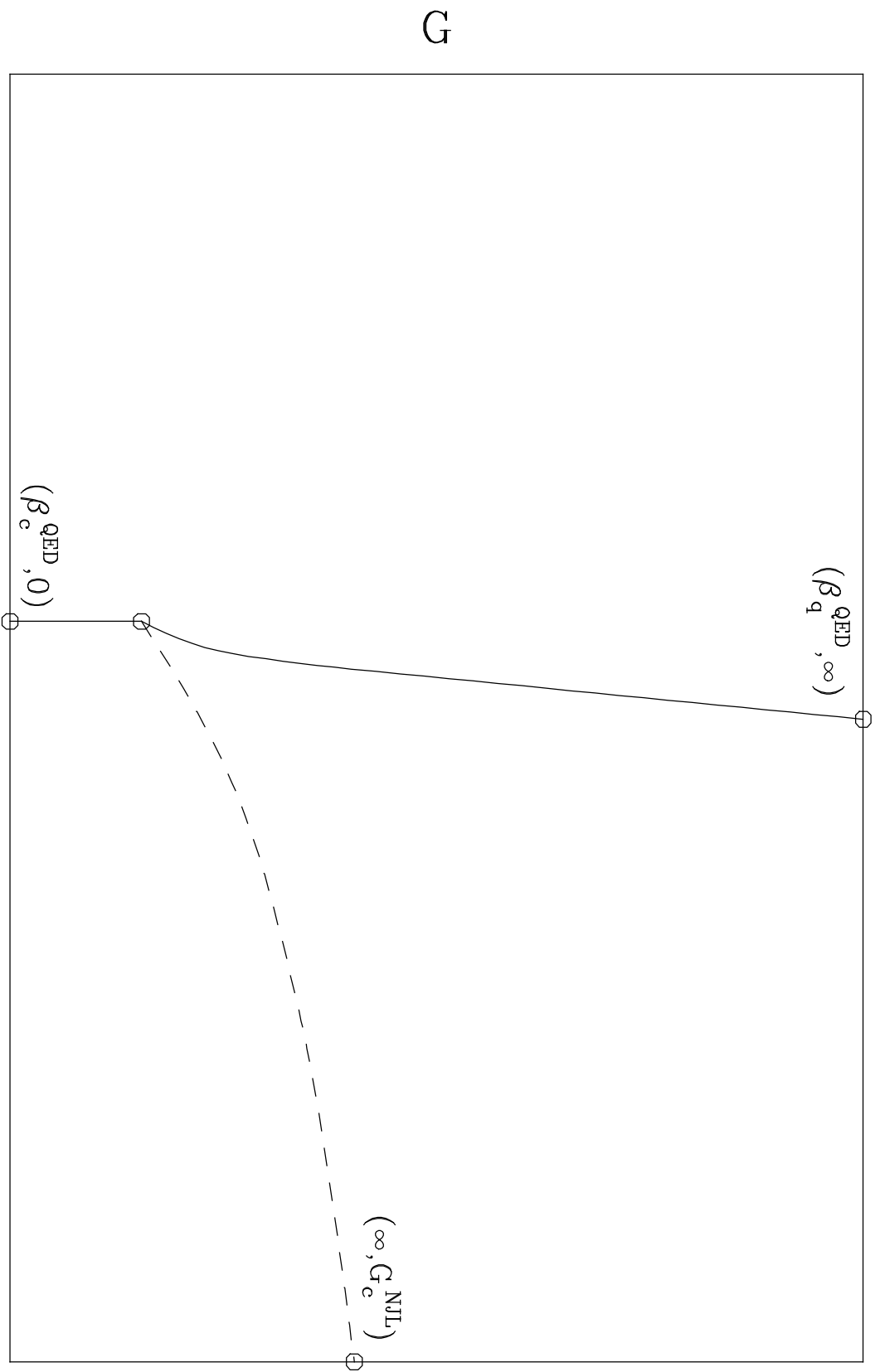
\end{document}